\documentclass{PoS}
\usepackage{epsfig}
\usepackage{cite}

\title{Topological susceptibility in the SU(3) random vortex
world-surface model }

\ShortTitle{Topological susceptibility in the SU(3) random vortex
world-surface model }

\author{\speaker{Michael Engelhardt}
\thanks{Supported by the U.S.~DOE under grant DE-FG02-96ER40965.}
\\
        Department of Physics, New Mexico State University, 
        Las Cruces, NM 88003, USA\\
        E-mail: \email{engel@nmsu.edu}}

\abstract{The topological charge is constructed for SU(3) center vortex
world-surfaces composed of elementary squares on a hypercubic lattice.
In distinction to the SU(2) case investigated previously, it is necessary
to devise a proper treatment of the color structure at vortex branchings,
which arise in the SU(3) case, but not for SU(2). The construction is used
to evaluate the topological susceptibility in the random vortex
world-surface model of infrared Yang-Mills dynamics. Results for the
topological susceptibility are reported as a function of temperature,
including both the confined as well as the deconfined phase.}

\FullConference{The XXVI International Symposium on Lattice Field Theory \\
                 July 14 - 19, 2008\\
                 Williamsburg, Virginia, USA}

\begin{document}

\section{Introduction}
The random vortex world-surface model
\cite{m1,m2,m3,su3conf,su3bary,su3freee,su4,sp2}
aims to describe the gluonic
dynamics of the strong interaction vacuum on the basis of center
vortex degrees of freedom. These vortices are closed tubes of
quantized chromomagnetic flux which spontaneously condense in the
vacuum, giving rise to the nonperturbative phenomena which characterize
the infrared sector of strong interaction physics, namely, confinement,
the spontaneous breaking of chiral symmetry and the axial $U_A (1)$
anomaly.

The random vortex world-surface model, by exploring the consequences of a
simplified dynamics governing a reduced, infrared effective set of (vortex)
degrees of freedom, complements studies of center vortices extracted from
lattice gauge configurations \cite{jg1,jg2,df1,df2,per,rb,hoellw}.
Such studies employ appropriate gauge fixing and projection procedures
\cite{jg1,jg2,df2} to determine the vortex content of gauge configurations;
having isolated the vortices, one can investigate to what extent, e.g.,
confinement is caused specifically by those vortices. Conversely, one can
ask whether any of the characteristic strong interaction phenomena
persist if vortices are removed from the lattice configurations. These
investigations have yielded strong support for the vortex picture of the
strong interaction vacuum; pars pro toto, results include vortex dominance
of the string tension \cite{jg1,jg2,df2}, the absence of topological charge
and a chiral condensate when vortices are removed \cite{df1,df2}, and a
picture of the deconfining phase transition in terms of vortex percolation
properties \cite{per}. Further details, including an outlook on the relation
of the vortex picture to other models of the strong interaction vacuum, can
be found in the reviews \cite{jg3,lat04}.

On the other hand, the simplified infrared effective description provided
by the random vortex world-surface model permits the exploration of a
wider variety of settings. For the simplest case of $SU(2)$ color, studied
initially, the model was shown to exhibit both a low-temperature confining
phase as well as a high-temperature deconfined phase \cite{m1}, separated
by a second-order phase transition \cite{su3conf}. The predictions of the
model for the spatial string tension in the deconfined phase \cite{m1},
the topological susceptibility \cite{m2} and the (quenched) chiral
condensate \cite{m3} quantitatively match corresponding $SU(2)$ lattice
Yang-Mills results. Building on this initial progress, the random vortex
world-surface model was extended to the case of $SU(3)$ color. Studies of
the confinement properties yielded a weakly first-order deconfinement
phase transition \cite{su3conf}, also seen in the vortex free energy
\cite{su3freee}, which represents an alternative order parameter for
confinement. Furthermore, a Y law for the baryonic static potential
was observed \cite{su3bary}. Spurred by recent investigations of
Yang-Mills theories with a wider variety of gauge groups, aiming
at a better understanding of possible confinement mechanisms \cite{spn},
the confinement properties of the random vortex world-surface model were
subsequently also studied for $SU(4)$ color \cite{su4} and $Sp(2)$ color
\cite{sp2}. These studies showed that the vortex picture can accomodate
such diverse color symmetries, while indicating the limitations of the
very simple effective dynamics which had proven adequate in the $SU(2)$
and $SU(3)$ cases.

After these detours into investigations of other gauge groups, the present
work returns to the $SU(3)$ case, widening the scope from the
confinement properties studied previously, cf.~above, to the topological
properties. Specifically, the topological charge is constructed for $SU(3)$
model vortex world-surfaces, where some additional complications arise
compared to the $SU(2)$ case. On this basis, the topological susceptibility,
which determines the anomalously high mass of the $\eta^{\prime } $ meson,
is evaluated as a function of temperature.

\section{SU(3) center vortex color structure}
\label{colorsec}
\begin{figure}
\vspace{-0.3cm}
\centerline{
\epsfig{file=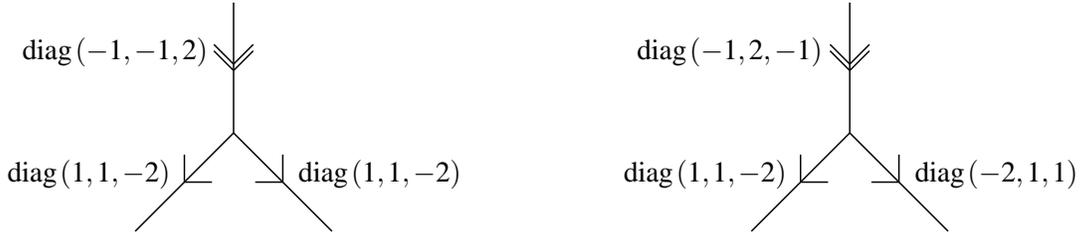,width=4cm}
\hspace{4cm}
\epsfig{file=branch.eps,width=4cm}
}
\vspace{-3.3cm}
\hspace{0.6cm} diag$\, (-1,-1,2)$ \hspace{5.5cm} diag$\, (-1,2,-1)$
\vspace{1.1cm}

\hspace{0.4cm} diag$\, (1,1,-2)$ \hspace{1.5cm} diag$\, (1,1,-2)$
\hspace{1.95cm} diag$\, (1,1,-2)$ \hspace{1.5cm} diag$\, (-2,1,1)$
\vspace{0.7cm}
\caption{Vortex color structure at branchings for a minimal choice of
the set of allowed color generators $T$ (left) and a non-minimal one (right);
cf.~main text.}
\label{branchfig}
\end{figure}

Center vortices are closed tubes of quantized chromomagnetic flux in
three space dimensions; correspondingly, they are described by (thickened)
closed two-dimensional world-surfaces in four space-time dimensions. The
quantization of flux is determined by the center of the gauge group. If
one evaluates a Wilson loop $W_C $ corresponding to a path $C$ encircling
an $SU(3)$ vortex, one obtains one of the two nontrivial center elements
of $SU(3)$,
\begin{equation}
W_C = \frac{1}{3} \, \mbox{Tr} \, {\cal P} \exp \left(
i\oint_{C} A_{\mu } dx_{\mu } \right) = \exp (\pm 2\pi i/3) \ .
\label{fluxquant}
\end{equation}
The two center elements in (\ref{fluxquant}) are complex conjugates;
thus, there is only one type of vortex flux, the two possible
orientations of which generate the two signs in (\ref{fluxquant}).
This color structure nevertheless admits vortex branching, in
contradistinction to the $SU(2)$ case; a vortex associated with a
center phase $\exp (-2\pi i/3) = \exp (4\pi i/3)$ can branch into
two vortices which are each associated with the phase $\exp (2\pi i/3)$.
This respects the Bianchi constraint (continuity of flux modulo Abelian
magnetic monopoles). The presence of branchings introduces an additional
complication for the evaluation of the topological charge of model vortex
world-surfaces, an issue which deserves further elaboration: Evaluating
the topological charge $Q$ via
\begin{equation}
Q=\frac{1}{32\pi^{2} } \int d^4 x \, \epsilon_{\mu \nu \lambda \tau } \
\mbox{Tr} \ F_{\mu \nu } F_{\lambda \tau }
\label{topcharge}
\end{equation}
implies that a more specific description of the vortex gauge field must be
given than just the Wilson loops (\ref{fluxquant}). As a first step, it
is useful to cast vortices in an Abelian gauge, i.e., the gauge field
$A_{\mu } $ and the field strength $F_{\mu \nu } $ are diagonal in color
space. This, however, does not yet completely specify the color structure.
One can further fix the gauge by allowing the vortex gauge field to be
proportional only to color matrices $T$ from a certain set. Consider the
following (minimal) choice,
\begin{equation}
T \in \{ \pm \, \mbox{diag} \, (1,1,-2) \} \ ,
\label{minchoice}
\end{equation}
so that there is one, and only one, generator for each of the possible vortex
fluxes,
\begin{equation}
W=(1/3) \, \mbox{Tr} \, \exp (2\pi iT/3) = \exp (\pm 2\pi i/3) \ .
\end{equation}
Consider now what happens at a branching of a vortex associated with
$T=\mbox{diag} \, (-1,-1,2)$ into two vortices each associated with
$T=\mbox{diag} \, (1,1,-2)$, cf.~Fig.~\ref{branchfig} (left).
Flux at the branching is only continuous modulo $2\pi $ in each color
component, since
\begin{equation}
\frac{2\pi }{3} \, \mbox{diag} \, (-1,-1,2) \ = \
2\cdot \frac{2\pi }{3} \, \mbox{diag} \, (1,1,-2) \ - \
2\pi \, \mbox{diag} \, (1,1,-2) \ .
\end{equation}
Therefore, there is necessarily a source or sink of Abelian magnetic flux,
i.e., an Abelian magnetic monopole associated with the branching. As will
become more clear further below, this introduces an ambiguity into the
evaluation of the topological charge which can only be remedied by deforming
the monopole world-line away from the branching. The minimal choice of color
matrices considered above is clearly not flexible enough to allow for such
a deformation. Consider, therefore, a non-minimal choice \cite{cw1} such as
\begin{equation}
T \in \left\{ \pm \, \mbox{diag} \, (1,1,-2) \, , \ 
\pm \, \mbox{diag} \, (1,-2,1) \, , \ \pm \, \mbox{diag} \, (-2,1,1) \,
\right\} \ .
\label{nonminchoice}
\end{equation}
In such a gauge, one has the freedom to disassociate monopoles from
branchings, cf.~Fig.~\ref{branchfig} (right); there,
\begin{equation}
\frac{2\pi }{3} \, \mbox{diag} \, (-1,2,-1) \ = \
\frac{2\pi }{3} \, \mbox{diag} \, (1,1,-2) \ + \
\frac{2\pi }{3} \, \mbox{diag} \, (-2,1,1) \ ,
\end{equation}
without the presence of any monopoles at the branching. This property
makes the non-minimal description (\ref{nonminchoice}) suited for an
unambiguous evaluation of the vortex topological charge, as opposed to the
minimal description (\ref{minchoice}). Vortex world-surfaces in the
following are thus described as consisting of patches which are each
associated with a specific color direction $T$ from the set
(\ref{nonminchoice}).

\section{Center vortex topological charge}
\label{chargesec}
Vortex topological charge is generated by two geometrical features, namely,
vortex world-surface intersection points and vortex world-surface
writhe\footnote{This is easily understood by noting that, in three
dimensions, a static magnetic flux in the 3-direction is associated
with the 3-component of the magnetic field, $B_3 $, and therefore with
the field strength component $F_{12} $. Expanding to space-time, this
means that a vortex world-surface running in the 3-4 directions is
associated with $F_{12} $ and, conversely, a world-surface running in
the 1-2 directions is associated with $F_{34} $. If the two intersect,
there is a point in space-time which simultaneously supports nonvanishing
$F_{12} $ and $F_{34} $, and at which topological charge density is
therefore generated according to (\ref{topcharge}). Similarly, if a
single vortex world-surface writhes in a way which explores all four
space-time dimensions, topological charge density is generated.}
\cite{m2,cw1,contvort,bruck,jordan}. As described further in section
\ref{modelsec}, vortex world-surfaces will, for practical purposes, be
modeled as consisting of elementary squares on a hypercubic lattice.
While for continuum surfaces, topological charge due to writhe is
distributed continuously on the surfaces \cite{bruck}, on the
aforementioned hypercubic model surfaces, topological charge due to writhe
is concentrated onto lattice sites (as is, of course, topological charge
due to vortex intersection points). Thus, at first sight, it would seem
that the topological charge of vortex surfaces composed of elementary
squares can be evaluated simply by considering all lattice sites in turn,
at each site finding all pairs of mutually orthogonal vortex squares
attached to the site, and weighting their contribution to the topological
charge appropriately with $\mbox{Tr} \, (T_1 T_2 )$, where $T_1 $ and
$T_2 $ are the color directions, cf.~(\ref{nonminchoice}), associated
with the two squares making up the pair in question. However, the
hypercubic description introduces two types of ambiguities which must
be resolved before the topological charge can be evaluated in this manner.

\begin{figure}
\vspace{-0.3cm}
\centerline{
\epsfig{file=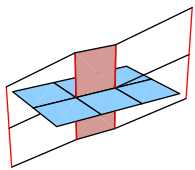,width=2.7cm}
}
\vspace{-0.8cm}
\caption{Composing vortex world-surfaces from elementary squares on a
hypercubic lattice introduces ambiguities, such as shared links between
surfaces, which are absent in generic continuum surface configurations.}
\label{intline}
\end{figure}

First, whereas continuum surfaces generically intersect at points in
four dimensions, random hypercubic lattice surfaces share whole line
segments with a finite probability, e.g., as displayed in Fig.~\ref{intline}.
Such features are artefacts due to the discrete set of directions
available to the surfaces on a lattice. They complicate the identification
of intersection points. When whole links are shared between surfaces, in
general there is no unambiguous assignment of the attached elementary
squares to two distinct surfaces; as a consequence, the question of whether
the surfaces are actually intersecting or merely touching becomes at best
nonlocal, and in general ambiguous. To resolve such ambiguities, the
hypercubic surfaces are transferred to a finer lattice, and small
deformations are carried out until there are no lattice links left to which
four or more vortex squares are attached (links with three attached squares
of course constitute bona fide branchings). The deformations are such that
the deformed surfaces never deviate from the original ones by more than
one-half of the original lattice spacing; therefore, on the scale of the
cutoff of the infrared effective vortex dynamics, the original and the
deformed surfaces are indistinguishable. In particular, all Wilson loops
defined on the original configurations remain unchanged. Nevertheless,
this procedure may introduce additional spurious topological charge
lumps on the scale of the finer lattice, and the topological susceptibility
may therefore be overestimated as a result. The corresponding systematic
downward uncertainty associated with the measured topological susceptibility,
cf.~Fig.~\ref{chifig} below, is quantified by scaling with the appropriate
power of the vortex density, which is expected to roughly track the
topological density, and which increases appreciably in the course of the
deformation procedure. This issue constitutes the chief source of
uncertainty in the results of this work.

The second ambiguity present in hypercubic surface configurations was
already alluded to in section~\ref{colorsec}. If Abelian magnetic monopoles
are rigidly tied to branchings, their world-lines necessarily run through
lattice sites. On the other hand, topological charge density is concentrated
onto the sites in the hypercubic description. This coincidence of the two
also creates an ambiguity; at monopoles, the color direction of the vortex
surface changes, and consequently, by having a monopole precisely intersect
a site carrying topological charge density, the latter can be arbitrarily
modified at that site. Such precise coincidence is again an artefact of
the hypercubic description, and this ambiguity is resolved by deforming
all monopole lines such that they never intersect sites carrying
topological charge. This is possible using the vortex color structure
defined by (\ref{nonminchoice}).

\section{Random vortex world-surface model}
\label{modelsec}

\begin{figure}
\vspace{-0.3cm}
\centerline{
\epsfig{file=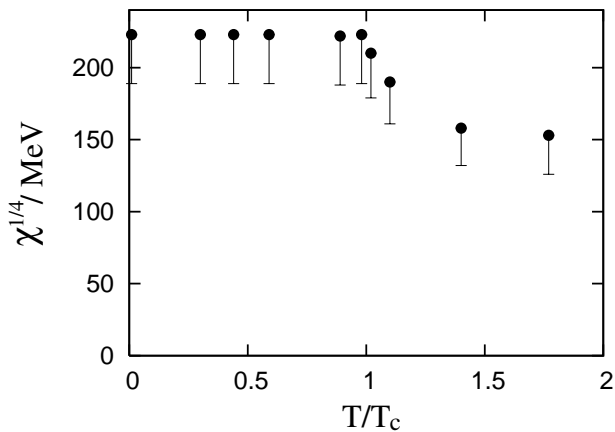,width=6cm,angle=-90}
}
\caption{Fourth root of the topological susceptibility measured in the
$SU(3)$ random vortex world-surface model, as a function of temperature.
Uncertainties due to statistical fluctuations of the random surface ensemble
are smaller than the filled circle symbols displaying the measured values;
the downward uncertainty shown is a systematic one, discussed in detail
in section~3.}
\label{chifig}
\end{figure}

The assumptions underlying the random vortex world-surface model have been
discussed in detail elsewhere \cite{m1,su3conf}; here, a short synopsis shall
suffice. Vortex world-surfaces are modeled by composing them of elementary
squares on a hypercubic lattice. An ensemble of such surfaces is generated
using Monte Carlo methods, respecting the (Bianchi) constraint that vortex
surfaces be closed at each update. The spacing of the lattice is a fixed
physical quantity, representing the ultraviolet cutoff of this infrared
effective model. Physically, it mimics vortex thickness; e.g., two parallel
thick vortices must remain at a minimal distance in order to still be
resolved as distinct. If one sets the scale by using the string tension
value $\sigma = (440\, \mbox{MeV} )^{2} $, the lattice spacing is
$0.39\, \mbox{fm} $. Finally, the vortex ensemble is governed by a
world-surface curvature action, symbolically,
\vspace{-0.4cm}

\begin{figure}[h]
\centerline{\hspace{0.1cm}
\epsfig{file=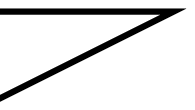,width=1.2cm} }
\end{figure}
\vspace{-1.8cm}

\begin{equation}
\hspace{-3cm} S\ \ = \ \ c \ \times
\end{equation}
\vspace{-0.4cm}

\noindent
i.e., each instance of two vortex squares sharing a link, but not lying in
the same plane costs an action increment $c$. The lone dimensionless
parameter $c$ of the model is fixed to $c=0.21$ by reproducing the
relation between the deconfinement temperature and the zero-temperature
string tension known from $SU(3)$ lattice Yang-Mills theory \cite{boyd},
$T_c /\sqrt{\sigma } =0.63$.

Generating a random vortex world-surface ensemble in this manner, and
determining the topological charge of each configuration as indicated in
section~\ref{chargesec} leads to the result for the topological
susceptibility $\chi = \langle Q^2 \rangle /V$ (where $V$ denotes the
four-volume) depicted in Fig.~\ref{chifig}; the temperature is varied as
usual by changing the extent of the Euclidean time dimension of the lattice,
and the downward uncertainty of the measurement was estimated according to
the discussion in section~\ref{chargesec}.

\section{Conclusions}
The raw data for the topological susceptibility collected in the $SU(3)$
random vortex world-surface model, cf.~Fig.~\ref{chifig}, are somewhat
higher than measurements in $SU(3)$ lattice Yang-Mills theory \cite{panag}
(cf.~in particular Table 1 therein), which roughly indicate a value of
around 190 MeV for the fourth root of the susceptibility at zero temperature,
with values as high as 213 MeV and as low as 168 MeV reported; for finite
temperatures, cf., e.g., \cite{alles}. However, there is a substantial
systematic downward uncertainty inherent in the measurement of the vortex
model topological susceptibility carried out in the present work. This
uncertainty is a consequence of ambiguities introduced into the
world-surface configurations by the hypercubic description,
cf.~section~\ref{chargesec}, and of the procedure employed to
resolve these ambiguities. Taking this uncertainty into account
leaves room for the possibility that the topological susceptibility
of the $SU(3)$ random vortex world-surface ensemble is consistent with
$SU(3)$ Yang-Mills theory. It should be stressed that the aforementioned
ambiguities are not intrinsic to the vortex picture, but result from the
specific hypercubic realization of the vortex world-surfaces adopted here
for practical purposes. An alternative model description which removes
the associated uncertainty could, e.g., be achieved by constructing the
world-surfaces as random triangulations. In this way, by letting surfaces
extend into arbitrary directions in four-dimensional space-time, the
ambiguities in question are avoided. On the other hand, such a
description is technically considerably more involved, e.g., in terms
of generating a grand canonical ensemble of surfaces, or also in terms
of the book-keeping of vortex locations.

\end{document}